\documentclass[aps,prl,twocolumn,superscriptaddress]{revtex4}

\usepackage{graphicx}
\usepackage{amsmath}

\newcommand{\eq}[1]{(\ref{#1})}
\newcommand{\kb}{k_{\rm B}}
\newcommand \lan {\langle}
\newcommand \ran {\rangle}

\begin{document}

\title{Fragility of the glassy state
of a directed polymer in random media:\\  a replica approach
}
\author{Marta Sales\footnote{msales@ffn.ub.es}
}
\affiliation{
Departament de F\'{\i}sica Fonamental, 
Facultat de F\'{\i}sica, Universitat de Barcelona
Diagonal 647, 08028 Barcelona, Spain
}
\author{
H. Yoshino\footnote{yoshino@ess.sci.osaka-u.ac.jp}
}
\affiliation{
Department of Earth and Space Science, Faculty of Science,
Osaka University, Toyonaka, 560-0043 Osaka, Japan
}


\begin{abstract}
We study the stability of the glassy equilibrium state of a directed 
polymer in random media against weak perturbations
including the intriguing problem temperature-chaos.
The statistical correlation of two real replica
systems under perturbations, such as small difference of 
temperature, is studied using the mapping to a quantum 
problem of interacting bosons.
Explicit replica symmetry breaking takes place 
being induced by the infinitesimally weak perturbations 
in the $n\to 0$ limit followed by the thermodynamic limit
which leads to complete decorrelation of the two real replicas.
The overlap lengths anticipated by phenomenological scaling arguments 
are retrieved as a correlation length associated with this 
explicit replica symmetry breaking.

\end{abstract}

\maketitle

A standing problem of glassy systems with randomness and frustration
is the possible instability of the glassy frozen states against 
infinitesimally weak perturbations such as an infinitesimal 
change of temperatures. A class of phenomenological scaling 
theories started first in the context of spin-glasses \cite{droplet}
suggests that the equilibrium glassy state dramatically changes
beyond the so called overlap length.
Unfortunately, the validity of the prediction has not been proven
explicitly except for some Migdal Kadanoff 
real space renormalization-group (MKRG) studies \cite{MKRG}.
The problem remains very far from being resolved especially
concerning the problem of temperature-chaos, i.~e. the sensitivity
to small temperature changes \cite{numerical,MFT-chaos}. 

The problem of temperature-chaos is now of great interest since it is 
considered as a possible mechanism of the
rejuvenation (restart of aging)  found in experiments such as
spin-glasses \cite{JVHBN} and polymer glasses \cite{polymer-glass}.
Concerning the memory 
effect which is simultaneously observed
in experiments, recently a novel dynamical memory 
was found in a coarsening model under explicit cycling 
of target equilibrium states \cite{YLB00}.

We consider a much simpler model, the directed 
polymer in random media (DPRM) in 1+1 dimension \cite{HH85,dprm-review}
which allows a much better access to the problem. 
It is known to be in the frozen phase at all finite temperatures 
in the sense that its scaling properties are always governed 
by the $T=0$ glassy fixed point.
In spite of its simplicity compared to spin-glass models, 
DPRM is believed to possess many of the subtle properties 
of glassy systems. Indeed the anomalous sensitivity 
of the glassy states to weak perturbations 
are predicted by scaling arguments which are partially supported
by numerical studies \cite{Z87,FV88,M90,S91,FH91,HH93}.
In the present paper, we present a unified 
analytical approach to deal with a variety of perturbations based 
on the concept of explicit replica symmetry breaking \cite{PV89,P90}
and confirm analytically the predictions of the scaling arguments 
for the first time.

DPRM belongs to the class of elastic manifolds 
in random media \cite{manifold-theory} so that 
its statistical mechanical properties
are potentially relevant for a broad class 
of physical systems of much interest.
The latter includes the domain walls of 
ferromagnets \cite{MDW} with weak bond 
randomness and the flux lines in type-II superconductors 
with randomly distributed point like pinning centers
\cite{bolle}, CDW  and vortex lattice systems 
with weak random-periodic pinnings \cite{CDW,blatt}.

We study the model described by the following Hamiltonian,
\begin{equation}
  H[V, h, \phi]   =   \int_{0}^{L} dz 
 \left [\frac{\kappa}{2}
  \left (\frac{d \phi(z)}{d z} \right)^{2}
+   V(\phi(z),z) 
\right].
\label{hamiltonian}
\end{equation}
The scalar field $\phi$  represents the displacement of the 
elastic object at point $z$ in a 1-dimensional internal space
of size $L$.  
The 1st term in the Hamiltonian is the elastic energy, $\kappa$ 
being the elastic constant. The random pinning medium is modeled 
as a quenched random potential $V(\phi,z)$ with zero mean
$\overline{V(\phi,z)}=0$ 
and short-ranged spatial correlation, 
$\overline{V\phi,z)V(\phi',z')} =
2D\delta(\phi-\phi')\delta(z-z').$ Here and in the following
$\overline{\cdots}$ stands for the average over different realizations
of the random potential.

Our starting point is a 
system of two real replicas, say A and B, 
whose configurations  $\phi_{A}(z)$  and $\phi_{B}(z)$ are subjected to
exactly the same random potential and temperature. 
Second, we apply small perturbations to them. In the present letter,
we consider 
\begin{itemize}
\item[i)]{\it Temperature difference} \cite{FH91}:
slightly different temperatures $T_{A}=T + \delta T$  $T_{B}=T - \delta T$ 
for A and B respectively with $\delta T/T \ll 1$
\item[ii)] {\it Decorrelation of random potential} \cite{Z87,FV88}: the random potential
of B is made from that of A
as $V_{B}=(V_{A}+\delta V')/\sqrt{1+\delta^{2}}$
where $|\delta| \ll 1 $  and $V'$ follows the same Gaussian distribution
as $V$. Then $\overline{V_{G}(\phi,z)}=0$ 
and $\overline{V_{G}\phi,z)V_{G'}(\phi',z')} =
2D_{G G'}\delta(\phi-\phi')\delta(z-z')$
with $D_{AA}=D_{BB}=D$ and $D_{AB}=D/\sqrt{1+\delta^{2}} < D$.
\item[iii)]  {\it Explicit short-ranged repulsive 
coupling} \cite{P90,M90}: A and B replicas are subjected to
explicit repulsive short-ranged interaction
$ \epsilon \int_{0}^{L}dz \delta (\phi_{A}(z)-\phi_{B}(z))$
with $0 < \epsilon \ll 1$.
\item[iv)] {\it  Tilt field} \cite{M90}: A and B replicas are subjected
to a tilting field of opposite sign
$
-h \phi_A(L) +h \phi_B(L)
$
with $h \ll 1$.
\end{itemize}


To study the consequence of perturbations, we analyze disorder average 
of partition function $\overline{Z^{n}_{A+B}(L)}$ 
in which both real replicas $G=A,B$ 
are further replicated into $n$-replicas $\alpha=1,\ldots,n$.
As noticed by Kardar \cite{K87}, if an analytical continuation 
for $n \to 0$ is  possible, such a  replicated partition function 
can be identified as a generator of cumulant correlation functions
of sample-to-sample fluctuations of free-energies \cite{K96}. 
\begin{eqnarray}
 \ln \overline{Z_{A+B}^{n}(L)}
&=& n \overline{[-\beta_A F_{A}(L)-\beta_B F_{B}(L)]} \nonumber \\ 
&+& \frac{n^{2}}{2} \overline{[-\beta_{A} F_{A}(L)-\beta_{B} F_{B}(L)]_{c}^2}
+ \ldots~~~~~
\label{eq5}
\end{eqnarray}
where $[\ldots]^{p}_{c}$ stands for $p$-th cumulant correlation functions of 
the total  free-energies $-\beta_A F_{A}(L)-\beta_B F_{B}(L)$ 
with $F_{G}(L)$ and $\beta_{G}$ being free-energy and inverse temperature of
each subsystem G= A, B.

Then if there is a complete change of the free-energy landscape due to
infinitesimally weak perturbations of strength $\delta \ll 1$, 
the statistical correlations between A and B 
should be lost asymptotically in the triple limits :
$n \to 0$ followed by $L\to\infty$  and finally $\delta \to 0$. 
Here the order of the limits is crucial.
Then the total partition function $\overline{Z^{n}_{A+B}(\delta,L)}$ 
under perturbation
should reduce to a product of the partition functions 
of the sub-systems $\overline{Z^{n}_{A}(L)}$  
and $\overline{Z^{n}_{B}(L)}$ for A and B respectively,
\begin{equation}
\lim_{\delta \to 0}\lim_{L \to \infty}\lim_{n \to
0}\overline{Z^{2n}_{A+B}}(\delta,L)
=\overline{Z^{n}_{A}(L)}\times \overline{Z^{n}_{B}(L)}
\label{eq-def-chaos}
\end{equation}
We use this factorization as the definition of the decorrelation of
the free-energy landscape ({\it chaos}).

To be specific, we consider that one end of all replicas is fixed
at $x=0$ and the other end fixed at $\{x_{G,\alpha}\}$. 
The partition function of the total system
can be expressed by a path integral
over all possible configurations of $2 \times n$ replicas,
\begin{eqnarray}
&& \overline{Z_{A+B}^{n}(\{x_{G,\alpha}\},L)} =
\int  \prod_{\substack{G=A,B\\ \alpha=1,\ldots,n}}{\cal D}\phi_{G,\alpha}
e^{-  S_{A+B}[\phi_{G,\alpha}]}~~~ ~~
\label{partition-function-start}
\end{eqnarray}
with a certain effective action $S_{A+B}[\phi_{G,\alpha}]$.
Then in turn, the partition function can be generated as
a path integral of a quantum system in imaginary time
with a Schr\"odinger equation,
\begin{equation}
-\frac{d}{dL}\overline{Z^{n}_{A+B}(\{x_{G,\alpha}\},L)}=
{\cal H}\overline{Z^{n}_{A+B}(\{x_{G,\alpha}\},L)}.
\end{equation}
with the Schr\"odinger operator
\begin{eqnarray}
&& {\cal H}=-\sum_{G,\alpha} \frac{\kb T_{G}}{2\kappa} \frac{\partial^{2}}{\partial x_{G,\alpha}^{2}} \nonumber \\
&& -  \sum_{\substack{(G,\alpha),\\ (G',\beta)}}
\frac{D_{GG'}}{(\kb T_{G})(\kb T_{G'})}
\delta (\phi_{G,\alpha}(z)-\phi_{G',\beta}(z)) \nonumber \\
&& +\epsilon\sum_{\alpha}
\delta (\phi_{A,\alpha}(z)-\phi_{B,\alpha}(z)) \nonumber \\
&& - \frac{h}{\kb T} \sum_{\alpha} \frac{\partial}{\partial x_{A,\alpha}}
+ \frac{h}{\kb T} \sum_{\alpha} \frac{\partial}{\partial x_{B,\alpha}}~~.
\label{eq2}
\end{eqnarray}
Here we have two different kinds of bosons for A and B each of which
has $n$ particles. The 1st term  represents the kinetic energy.
The 2nd term stands for attractive short-ranged interactions between
the bosons where the sum is taken over all possible pairs of bosons.
The 3rd term is the explicit repulsive coupling.
The last term is due to the  tilt field
applied to A and B subsets in the opposite directions.

The Shcr\"odinger operator can be decomposed as
\begin{equation}
{\cal H}_{A+B}={\cal H}_{0}+  \delta {\cal H}
\end{equation}
where 
\begin{eqnarray}
{\cal H}_0=-\sum_{G,\alpha} \frac{\kb T}{2\kappa} \frac{\partial^{2}}{\partial x_{G,\alpha}^{2}}
-\frac{D}{(\kb T)^{2}}
\sum_{\substack{((G,\alpha),\\
(G',\beta))}}\delta (x_{G,\alpha}-x_{G',\beta})~~
\label{eq-H0}
\end{eqnarray}
is the Schr\"odinger operator in the absence of perturbations and 
$\delta {\cal H}$ is a perturbation term representing 
one of the specific perturbations i)-iv).
Note that the original operator ${\cal H}_{0}$
has a high replica (permutation) symmetry: it is symmetric under all possible 
permutations of 
bosons not only among A and B subsets  but among the whole set. 

Obviously, the perturbation term $\delta {\cal H}$
breaks the replica (bosonic) symmetry explicitly. 
Then the interesting question is whether such a term can 
break the symmetry in the thermodynamic limit even if it is
infinitesimally small. If a decorrelation of the free-energy (chaos)
takes place,  it should manifest itself as reduction of this
high symmetry, i.~e. replica symmetry breaking, such that
replica symmetry remains at most only within each subset A or B.
Our problem can be considered as a particular example
of the problem of explicit replica symmetry breaking proposed
by Parisi and Virasoro \cite{PV89}, who tried to 
give a prescription to define replica symmetry breaking 
in a sound thermodynamic sense by introducing infinitesimally weak
replica symmetry breaking terms. Indeed our approach is an 
extension of the work by Parisi \cite{P90} who implemented
the idea of explicit replica symmetry breaking in the context of DPRM.


The ground state of the Schr\"odinger operator \ref{eq-H0} can be obtained
via the Bethe ansatz \cite{b-ansatz}. It is a bound state involving all particles
described by the wavefunction
\begin{equation}
<\Psi_{\rm RS}|\{x_{G,\alpha}\}> 
\sim \exp \left(-\lambda\sum_{((G,\alpha),(G',\beta))}
|x_{\alpha,G}-x_{\beta,G'}|\right)
\label{eqf}
\end{equation}
with $\lambda=\kappa D/(\kb T)^3$. Since it is completely 
replica symmetric, we call this as RS (replica symmetric) state.
Following Parisi \cite{P90}, we consider also an excited state
\begin{equation}
<\Psi_{\rm RSB}| = <\Psi^{A}_{\rm RS}| \times <\Psi^{B}_{\rm RS}|
\qquad <\Psi^{B}_{\rm RS}|\Psi^{A}_{\rm RS}>=0.
\label{zero-overlap}
\end{equation}
where $<\Psi^{A}_{\rm RS}|$ and $<\Psi^{B}_{\rm RS}|$ 
are similar bound states described by the wavefunction like \eq{eqf}
but constituted by only A and B particles.  
Assuming that the centers of mass of 
the two bound states are infinitely far from each other 
so that their overlap is zero, such a state becomes 
also an eigenstate.
Since replica symmetry is reduced here, we refer to the
state as RSB (replica symmetry broken) state.

Let us consider the difference (gap) of the eigen value
of the RS ground state and the RSB excited state under the
perturbation
\begin{equation}
\Gamma (n)=\Gamma_{0}(n) +\delta \Gamma(n)
\label{gap}
\end{equation}
where
$
\Gamma_{0}(n)=\kb T/\kappa\:\lambda^2\: n^3 
$
is the original 'energy gap' and $\delta \Gamma(n)$ is the correction 
due to the perturbation. 
Assuming that the amplitudes of the perturbations are small
enough, one may compute the correction term at 1st order of perturbation as,
\begin{equation}
\delta \Gamma(n)=  \left (
\frac{\lan \Psi_{\rm RSB}|\delta {\cal H}|\Psi_{\rm RSB}\ran}{\lan
\Psi_{\rm RSB}|\Psi_{\rm RSB}\ran}
-\frac{\lan \Psi_{\rm RS}|\delta {\cal H}|\Psi_{\rm RS}\ran}{\lan
\Psi_{\rm RS}|\Psi_{\rm RS}\ran}
\right).
\label{gap-with-correction}
\end{equation}

As we outline later, the correction terms can be 
computed for the cases that we consider
and we find a generic form for the leading term 
\begin{equation}
-\delta \Gamma(n) \sim \Delta n^{p} \qquad 1 \leq p < 3
\label{gap-correction}
\end{equation}
with $\Delta>0$ being a constant which represents the strength
of perturbations and $p$ being a positive integer smaller than $3$.
We call the exponent $p$ as {\it order of perturbation}.
Here 'leading' term means a term with smallest power of $n$
which becomes most important in the $n \to 0$ limit that we consider.
Let us note here that we have chosen the detailed form of the
perturbations i)-iv) such that this gap \eq{gap-correction}
remains invariant under exchange $A \leftrightarrow B$.
Combining the results one finds,
\begin{equation}
\Gamma(n)=
\Gamma_{0}(n) \left [1- \left (\frac{n}{n^{*}(\Delta)} \right)^{-(3-p)}\right]
\label{eq-gap-crossing}
\end{equation}
with
\begin{equation}
n^{*}(\Delta)= \left(\frac{\Delta}{\lambda^2 \kb T/\kappa}\right)^{1/(3-p)}~~.
\label{eq-n-star}
\end{equation}
Let us recall that 
we have to take  the $n \to 0$ limit before $L \to \infty$. 
Since $p<3$ holds for all the cases we consider,
an arbitrarily small perturbation $\Delta$ will induce
a level crossing at $n^{*}(\Delta)$ below which the contribution of 
RSB excited state to the replicated partition function 
becomes larger than that of the 
original ground state (RS), i.~e. there is replica symmetry breaking.
The result \eq{eq-gap-crossing} 
strongly suggests that the factorization of the 
partition function \eq{eq-def-chaos},
i.~e.~complete change of the free-energy landscape (or chaos), 
takes place in the thermodynamic limit even with infinitesimally
small (but non-zero) perturbation of strength $\Delta \ll 1$,
\begin{equation}
\lim_{\Delta \to 0}\lim_{L \to \infty} \lim_{n/n* \to
0}\overline{Z^{2n}_{A+B}(\Delta,L)}
=\overline{Z^{n}_{A}(L)}\times \overline{Z^{n}_{B}(L)}\qquad
\mbox{($p < 3$)}.
\label{eq-decouple}
\end{equation}

Now let us further exploit from the above result 
to find a more physical picture. 
In the absence of perturbations, 
the logarithm of the replicated partition
function is known to have a functional form \cite{BO90,K96} 
$
\ln\overline{Z_{A+B}^{n}(\Delta,L)}= -\beta \overline{f} L (2n) +g(2nL^{1/3})
$
which yields $\overline{[-\beta F]^{p}_{c}} \propto L^{p\theta}$
with  $\theta$ being the exact stiffness exponent $\theta=1/3$.
On the other hand, \eq{eq-gap-crossing} implies $n/n^{*}$ is another
natural variable of the replicated partition function \cite{EK00}.

Combining the two, we conjecture the following scaling ansatz,
\begin{eqnarray}
 \ln\overline{Z_{A+B}^{2n}(L)}+\beta \overline{f} L (2n) 
&& =g(2nL^{1/3}, n/n^*)\nonumber \\
&& =g(2nL^{1/3}, L/L^*)~~~
\label{ansatz}
\end{eqnarray}
where $\overline{f}$ is the average free-energy per unit length.
Here we introduced a characteristic length $L^*$ defined as
\begin{equation}
L^*\sim(n^*)^{-3} \sim \Delta^{-3/(3-p)}
\label{eq-L-star}
\end{equation} 
The $n \to 0$ limit induces the thermodynamic
limit $L \to \infty$ if the variable $nL^{1/3}=x$ is fixed.
Then for a fixed $x$, two limiting behaviors are expected.
First the small length scale regime $L/L^* \ll 1$ corresponds
to $n/n^* \gg 1$ so that
the effect of perturbation will be small and the partition function
is essentially the same as that of the unperturbed system of 
$2 \times n$-replicas
$
g\left(x, L/L^{*}\to 0\right)  \simeq  g(2x).
$
Second the large length regime  $L/L^* \gg 1$ 
corresponds to $n/n^* \ll 1$. Thus 
the decoupling \eq{eq-decouple} implies
$
\tilde{g}(x,  L/L^{*} \to \infty)\simeq  2 \times g(x).
$

The above result implies the crossover length $L^*$ should 
correspond to the overlap length $L_{c}$ anticipated 
by the scaling argument. In the following we present the
outline of our results for the specific perturbations.
Details will be presented elsewhere \cite{full-paper}.\\
\vspace{2pt}\\
{\em Temperature and potential change}\\
In the case of i) {\it small temperature difference},
one finds by the perturbation calculation that 
the correction to the gap \eq{gap-correction} is characterized
by order $p=2$ with amplitude $\Delta \propto (\delta T/T)^{2}$.
Then using \eq{eq-n-star} and \eq{eq-L-star}, we find
the overlap length $L^* \sim (\delta T/T)^{-6}$.
Similarly,  the case of ii) 
{\it small decorrelation of random potential}
also amounts to the same order $p=2$ with $\Delta \sim \delta^{2}$.
Then one finds the overlap length 
$L^* \sim \delta^{-6}$. 
Quite remarkably, the overlap-length obtained here is in agreement
with the prediction of the scaling arguments \cite{FV88,FH91}.
Furthermore, it is interesting to note that the two perturbations 
which look very different at a first sight, turn out to be the 
same in the replica space.\\
\vspace{2pt}\\
{\em Explicit repulsive coupling}\\
The case of iii) {\it explicit short-ranged repulsive 
coupling} has been considered by Parisi \cite{P90} and the result
gives  order $p=1$ and amplitude $\Delta \sim \epsilon$. Thus one finds
the overlap-length $L^* \sim \epsilon^{-3/2}$ which
agrees with the prediction \cite{M90} based on a scaling argument.\\
\vspace{2pt}\\
{\em  Tilt field}\\
Finally we consider the case of iv)  tilt field.
In this case, the 1st order correction term \eq{gap-correction}
vanishes. Fortunately, we can proceed as the following. 
The Schr\"odinger operator with the tilt field can be rewritten 
into the fully symmetric form as the unperturbed one \eq{eq-H0}
by shifting the momenta,
$\partial/\partial x'_{A,\alpha}=
\partial/\partial x_{A,\alpha} - h/\kb T$ 
and 
$\partial/\partial x'_{B,\alpha}
=\partial/\partial x_{B,\alpha}+h/\kb T$.
Therefore the exact ground state under the  tilt field
can be obtained from the Bethe's wave-function \eq{eqf} 
by undoing the previous shifting of moments.
On the other hand, the unperturbed RSB wavefunction \eq{zero-overlap}
remains as a valid eigenstate under the  tilt field.
Combining these results, we obtain the correction to the gap due to 
the  tilt field as
$\delta \Gamma(n)=n h^2/\kappa\kb T$
from which we read off order $p=1$ and amplitude $\Delta \sim h^{2}$,
which yields the overlap  length $L^*\sim h^{-3}$.
The latter agrees with the prediction
\cite{M90} based on a scaling argument.\\
\vspace{2pt}\\
To summarize we have presented a unified analytical approach
to the problem of the sensitivity to perturbations of the equilibrium glassy states
of 1+1 dimensional DPRM based on the idea of explicit
replica symmetry breaking. Our result in replica space
suggests  that perturbations can be classified according to their order $p$ 
and the symmetries that are preserved. 
Therefore, apparently different perturbations 
can bring about the same  effects. The universal features 
of the crossover from the weakly perturbed regime $L/L^* \ll 1$
to the strongly perturbed regime  $L/L^* \gg 1$ can be examined
numerically \cite{full-paper}. The prediction of the scaling 
arguments is supported for 4 different kinds of perturbations. 
It may be promising to apply
the approach of the present paper
in the studies of other glassy systems including spin-glasses.\\
\vspace{2pt}\\
{\bf Acknowledgements}\\
We thank M. M\'{e}zard, J. P. Bouchaud and F. Ritort for useful discussions.
We thank Service de Physique de l'\'Etat Condens\'e, CEA Saclay
for financial supports and kind hospitality where major part 
of this work was done.
M. S. acknowledges the MEC of the Spanish government for grant AP98-36523875.

\end{document}